\documentclass[10pt,letter]{article}                                                                   
\usepackage{graphicx}
\newcommand{\ud}{\mathrm{d}}

\begin{document}                                                                                    
\thispagestyle{empty}                                                              
                                                                                   
\begin{center}                                                                     
\begin{tabular}{p{130mm}}                                                          
                                                                                   
\begin{center}                                                                     
{\bf\Large                                                                         
LOCALIZATION AND FUSION MODELING }\\
\vspace{5mm}  

{\bf\Large IN PLASMA PHYSICS.} \\                                                  
\vspace{5mm}                                                                       
                                                                                   
{\bf\Large  PART II: VLASOV-LIKE SYSTEMS. }\\                                         
\vspace{5mm}                                                                       
                                                                                   
{\bf\Large IMPORTANT REDUCTIONS}\\                                                      
                                                                                   
\vspace{1cm}                                                                       
                                                                                   
{\bf\Large Antonina N. Fedorova, Michael G. Zeitlin}                             
                                                                                   
\vspace{1cm}

{\bf\large\it                                                                      
IPME RAS, St.~Petersburg,                                                          
V.O. Bolshoj pr., 61, 199178, Russia}\\                                            
{\bf\large\it e-mail: zeitlin@math.ipme.ru}\\                                      
{\bf\large\it e-mail: anton@math.ipme.ru}\\                                        
{\bf\large\it http://www.ipme.ru/zeitlin.html}\\                                   
{\bf\large\it http://www.ipme.nw.ru/zeitlin.html}                                
                                                                                   
\end{center}                                                                       
                                                                                   
\vspace{2cm}                                                                       
                                                                                   
\begin{abstract}                                                                   
  The methods developed in the previous Part I are applied to a few important 
reductions of BBGKY hierarchy, namely, various examples of Vlasov-like systems. It is well 
known that they are important both for fusion modeling and for particular physical problems 
related to plasma/beam physics. As in part I we concentrate mostly on phenomena of 
localization and pattern formation.
\end{abstract}

\vspace{10mm}                                                                      
                                                                                   
\begin{center}                                                                     
{\large 
Two lectures presented at the Sixth Symposium on
Current Trends in International Fusion Research, Washington D.C., March, 2005,}\\ 

{\large             
edited by Emilio Panarella, NRC Reasearch Press, National Reasearch Council of Canada,            
Ottawa, Canada, 2006.}                                                     
\end{center}
                                                                       
\end{tabular}                                                                      
\end{center}

\newpage

\title{LOCALIZATION AND FUSION MODELING IN 
PLASMA PHYSICS. PART II: VLASOV-LIKE 
SYSTEMS. IMPORTANT REDUCTIONS\footnote{
Current Trends in International Fusion Research - Proceedings of the Sixth Symposium
Edited by Emilio Panarella. NRC Reasearch Press, National Reasearch Council of Canada, 
Ottawa, ON K1A 0R6 Canada, 2006.}}
\author{Antonina~ N. Fedorova and Michael~ G. Zeitlin \\
IPME RAS, Russian Academy of Sciences, \\
V.O. Bolshoj pr., 61, \\
199178, St. Petersburg, Russia \\
http://www.ipme.ru/zeitlin.html;\\
http://www.ipme.nw.ru/zeitlin.html 
}
\date{}
\maketitle
\thispagestyle{empty}

\begin{abstract}      
   The methods developed in the previous Part I are applied to a few important 
reductions of BBGKY hierarchy, namely, various examples of Vlasov-like systems. It is well 
known that they are important both for fusion modeling and for particular physical problems 
related to plasma/beam physics. As in part I we concentrate mostly on phenomena of 
localization and pattern formation. 
\end{abstract}

\section{INTRODUCTION: VLASOV-POISSON\\ SYSTEM}
\subsection{Description}
         In this part we present the applications of our approach based on variational 
multiresolution technique [1]-[6], considered in Part I [7], to the systems with collective type 
behaviour described by some forms of Vlasov-Poisson/Maxwell equations, some important 
reduction of general BBGKY hierarchy [8]. 
         Such approach may be useful in all models in which it is possible and reasonable to 
reduce all complicated problems related to statistical distributions to the problems described 
by the systems of nonlinear ordinary/partial differential/integral equations with or without 
some (functional) constraints. In periodic accelerators and transport systems at the high beam 
currents and charge densities the effects of the intense self-fields, which are produced by the 
beam space charge and currents, determinine (possible) equilibrium states, stability and 
transport properties according to underlying nonlinear 
dynamics. The dynamics of such space-charge dominated 
high brightness beam systems can provide the understanding of the 
instability phenomena such as emittance growth, mismatch, halo formation related to the
complicated behaviour of underlying hidden nonlinear modes outside of perturbative tori-like 
KAM regions [8].
         Our analysis based on the variational-wavelet approach allows to consider 
polynomial and rational type of nonlinearities. 
         In some sense in this particular case this approach is direct generalization of 
traditional nonlinear $\delta F$  approach [8] in which weighted Klimontovich representation
\begin{eqnarray}
\delta f_j=a_j\sum^{N_j}_{i=1}w_{ji}\delta(x-x_{ji})\delta(p-p_{ji})
\end{eqnarray} 
or self-similar decompostion like 
\begin{eqnarray}
\delta n_j=b_j\sum^{N_j}_{i=1}w_{ji} s(x-x_{ji}),
\end{eqnarray}
where  $s(x-x_{ji})$ is a shape function of distributing particles on the grids in 
configuration space, are replaced by powerful technique from local nonlinear harmonic 
analysis, based on underlying symmetries of functional space such as affine or more general. 
         The solution has the multiscale/multiresolution 
decomposition via nonlinear high-localized 
eigenmodes, which corresponds to the full multiresolution expansion in all 
underlying time/phase space scales. 
         Starting from Vlasov-Poisson equations, we consider the approach based on 
multiscale variational-wavelet formulation. We give the explicit representation for all 
dynamical variables in the base of compactly supported wavelets or nonlinear eigenmodes. 
Our solutions are parametrized by solutions of a number of reduced algebraical problems, one 
from which is nonlinear with the same degree of nonlinearity as initial problem and the others 
are the linear problems which correspond to the particular method of calculations inside 
concrete wavelet scheme. Because our approach started from variational formulation we can 
control evolution of instability on the pure algebraical level of reduced algebraical system of 
equations. This helps to control stability/unstability scenario of evolution in parameter space 
on pure algebraical level. In all these models numerical modeling demonstrates the 
appearance of coherent high-localized structures and as a result the stable patterns formation 
or unstable chaotic behaviour. 
         Analysis based on the non-linear Vlasov equations leads to more clear understanding 
of collective effects and nonlinear beam dynamics of high intensity beam propagation in 
periodic-focusing and uniform-focusing transport systems. We consider the following form of 
equations
\begin{eqnarray}
&&\Big\{\frac{\partial}{\partial s}+p_x\frac{\partial}{\partial x}+
             p_y\frac{\partial}{\partial y}-
\Big[k_x(s)x+\frac{\partial\psi}{\partial x}\Big]\frac{\partial}{\partial p_x}-\nonumber\\
&& \Big[k_y(s)y+\frac{\partial\psi}{\partial y}\Big]\frac{\partial}{\partial p_y}
  \Big\} f_b(x,y,p_x,p_y,s)=0, \\
&&\Big(\frac{\partial^2}{\partial x^2}+\frac{\partial^2}{\partial y^2}\Big)\psi=
-\frac{2\pi K_b}{N_b}\int \ud p_x \ud p_y f_b,\nonumber\\
&&\int\ud x\ud y\ud p_x\ud p_y f_b=N_b.\nonumber
\end{eqnarray} 		
The corresponding Hamiltonian for transverse single-particle motion is given by 
 \begin{eqnarray}
&& H(x,y,p_x,p_y,s)=\frac{1}{2}(p_x^2+p_y^2) 
                   +\frac{1}{2}[k_x(s)x^2 \\
 &&+k_y(s)y^2]+
    H_1(x,y,p_x,p_y,s)+\psi(x,y,s), \nonumber
\end{eqnarray}		
where $H_1$  is nonlinear (polynomial/rational) part of the full Hamiltonian and corresponding 
characteristic equations are: 
\begin{eqnarray}
\frac{\ud^2x}{\ud s^2}+k_x(s)x+\frac{\partial}{\partial x}\psi(x,y,s)&=&0,\\
\frac{\ud^2y}{\ud s^2}+k_y(s)y+\frac{\partial}{\partial y}\psi(x,y,s)&=&0.\nonumber
\end{eqnarray}
         
\subsection{Multiscale Representation}

         We obtain our multiscale/multiresolution representations for solutions of these 
equations via variational-wavelet approach. We decompose the solutions as 
\begin{eqnarray}
&&f_b(s,x,y,p_x,p_y)=\sum^\infty_{i=i_c}\oplus\delta^if(s,x,y,p_x,p_y),\\
&&\psi(s,x,y)=\sum^\infty_{j=j_c}\oplus\delta^j\psi(s,x,y),\nonumber\\
&&x(s)=\sum^\infty_{k=k_c}\oplus\delta^kx(s),\nonumber\\
&&y(s)=\sum^\infty_{\ell=\ell_c}\oplus\delta^\ell y(s),\nonumber
\end{eqnarray}		
where set 
\begin{equation}
(i_c,j_c,k_c,\ell_c)
\end{equation}
corresponds to the coarsest level of resolution $c$  in the full multiresolution 
decomposition [9]
\begin{equation}
V_c\subset V_{c+1}\subset V_{c+2}\subset\dots
\end{equation}
Introducing detail space $W_j$   as the orthonormal complement of  $V_j$ with respect to 
\begin{equation}
V_{j+1}: 
V_{j+1}=V_j\bigoplus W_j,
\end{equation}
         we have for 
\begin{equation}
f, \psi, x, y \subset L^2({\bf R})
\end{equation}

\begin{eqnarray}
L^2({\bf R})=\overline{V_c\displaystyle\bigoplus^\infty_{j=c} W_j}.
\end{eqnarray}

         In some sense it is some generalization of the old $\delta F$  approach [8]. Let $L$  be an 
arbitrary (non) linear differential/integral operator with matrix dimension $d$, which acts on 
some set of functions
\begin{eqnarray}
\Psi\equiv\Psi(s,x)=\Big(\Psi^1(s,x),\dots,\Psi^d(s,x)\Big),\ 
 s,x \in\Omega\subset{\bf R}^{n+1}
\end{eqnarray}
         from  $L^2(\Omega)$: 
\begin{equation}
L\Psi\equiv L(R(s,x),s,x)\Psi(s,x)=0,
\end{equation}
where  $x$ are the generalized space coordinates or phase space coordinates, and  $s$ is 
"time" coordinate. After some anzatzes the main reduced problem may be formulated as the 
system of ordinary differential equations
                                                               
\begin{eqnarray}\label{eq:pol0}                                
& & Q_i(f)\frac{\ud f_i}{\ud s}=P_i(f,s),\quad f=(f_1,..., f_n),\\
& &i=1,\dots,n, \quad                                                                        
 \max_i  deg \ P_i=p, \quad \max_i deg \  Q_i=q \nonumber                  
\end{eqnarray}

\noindent          or a set of such systems corresponding to each independent coordinate in phase 
space. They have the fixed initial (or boundary) conditions $f_i(0)$, where $P_i, Q_i$  are not more 
than polynomial functions of dynamical variables $f_j$  and have arbitrary dependence on time. 
As result we have the following reduced algebraic system of equations on the set of 
unknown coefficients  $\lambda_i^k$ of localized eigenmode expansion:
\begin{eqnarray}\label{eq:pol2}
L(Q_{ij},\lambda,\alpha_I)=M(P_{ij},\lambda,\beta_J),
\end{eqnarray}
where operators L and M are algebraization of RHS and LHS of initial problem and $\lambda$ 
  are unknowns of reduced system of algebraical equations (RSAE). After solution of RSAE 
(15) we determine the coefficients of wavelet expansion and therefore obtain the solution of 
our initial problem. It should be noted that if we consider only truncated expansion with $N$   
terms then we have the system of  $N\times n$ algebraic equations with degree
\begin{eqnarray}
\ell=max\{p,q\}
\end{eqnarray}
and the degree of this algebraic system coincides with degree of initial differential 
system. So, we have the solution of the initial nonlinear (rational) problem in the form

\begin{eqnarray}\label{eq:pol3}
f_i(s)=f_i(0)+\sum_{k=1}^N\lambda_i^k f_k(s),
\end{eqnarray}
where coefficients  $\lambda_i^k$ are the roots of the corresponding reduced algebraic 
(polynomial) problem RSAE. Consequently, we have a parametrization of solution of initial 
problem by the solution of reduced algebraic problem. The obtained solutions are given in 
this form, where  $f_k(t)$ are basis functions obtained via multiresolution expansions and 
represented by some compactly supported wavelets. As a result the solution of equations has 
the following multiscale/multiresolution decomposition via nonlinear high-localized 
eigenmodes, which corresponds to the full multiresolution expansion in all underlying scales 
starting from coarsest one. For
\begin{equation}
{\bf x}=(x,y,p_x,p_y)
\end{equation}
         we will have 
\begin{eqnarray}\label{eq:z}
\Psi(s,{\bf x})&=&\sum_{(i,j)\in Z^2}a_{ij}{\bf U}^i\otimes V^j(s,{\bf x}),\\
V^j(s)&=&V_N^{j,slow}(s)+\sum_{l\geq N}V^j_l(\omega_ls), \quad \omega_l\sim 2^l \nonumber\\
{\bf U}^i({\bf x})&=&{\bf U}_M^{i,slow}({\bf x})+
\sum_{m\geq M}{\bf U}^i_m(k_m{\bf x}), \quad k_m\sim 2^m, \nonumber
\end{eqnarray}

         These formulae give us 
expansion into the slow part $\Psi_{N,M}^{slow}$  and fast oscillating parts 
for arbitrary $N, M$. So, we may move from coarse scales of resolution to the finest one for 
obtaining more detailed information about our dynamical process. The first terms in the RHS 
correspond on the global level of function space decomposition to resolution space and the 
second ones to detail space. It should be noted that such representations give the best possible 
localization properties in the corresponding (phase)space/time coordinates. In contrast with 
other approaches this formulae do not use perturbation technique or linearization procedures. 
So, by using wavelet bases with their good (phase) space/time localization properties we can 
describe high-localized (coherent) structures in spatially-extended stochastic systems with 
collective behaviour. 
         Modeling demonstrates the appearance of stable patterns formation from high-
localized coherent structures or chaotic behaviour. On Fig. 1 we present contribution to the 
full expansion from coarsest level of decomposition. Fig. 2 shows the representations for full 
solutions, constructed from the first six scales (dilations) and demonstrates (meta) stable 
localized pattern formation in comparison with chaotic-like behaviour (Fig. 10, Part I) outside 
of KAM region. We can control the type of behaviour on the level of reduced algebraic 
system (15).

\section{RATE/RMS MODELS}
\subsection{Description}
         In this part we consider the applications of our technique based on the methods of 
local nonlinear harmonic analysis to nonlinear rms/rate equations for averaged quantities 
related to some particular case of nonlinear Vlasov-Maxwell equations. 
         Our starting point is a model and approach proposed by R. C. Davidson e.a. [8]. We 
consider electrostatic approximation for a thin beam. This approximation is a particular 
important case of the general reduction from statistical collective description based on Vlasov-
Maxwell equations to a finite number of ordinary differential equations for the second 
moments related quantities (beam radius and emittance). In our case these reduced rms/rate 
equations also contain some distribution averaged quantities besides the second moments, e.g. 
self-field energy of the beam particles. Such model is very efficient for analysis of many 
problems related to periodic focusing accelerators, e.g. heavy ion fusion and tritium 
production. So, we are interested in the understanding of collective properties, nonlinear 
dynamics and transport processes of intense non-neutral beams propagating through a periodic 
focusing field. Our approach allows to consider rational type of nonlinearities in rms/rate 
dynamical equations containing statistically averaged quantities also. 
         The solution has the multiscale/multiresolution 
decomposition via nonlinear high-localized 
eigenmodes (waveletons), which corresponds to the full multiresolution expansion 
in all underlying internal hidden scales. We may move from coarse scales of resolution to the 
finest one to obtain more detailed information about our dynamical process. In this way we 
give contribution to our full solution from each scale of resolution or each time/space scale or 
from each nonlinear eigenmode. 
         Starting from some electrostatic approximation of Vlasov-Maxwell system and 
rms/rate dynamical models we consider the approach based on variational-wavelet 
formulation. We give explicit representation for all dynamical variables in the bases of 
compactly supported wavelets or nonlinear eigenmodes. Our solutions are parametrized by the 
solutions of a number of reduced standard algebraic problems.

\subsection{Rate Equations}
         In thin-beam approximation with negligibly small spread in axial momentum for 
beam particles we have in Larmor frame the following 
electrostatic approximation for Vlasov-Maxwell equations [8]:
\begin{eqnarray}
\frac{\partial F}{\partial s}+x'\frac{\partial F}{\partial x}+
y'\frac{\partial F}{\partial y}-\Big(k(s)x+\frac{\partial\psi}{\partial x}\Big)
\frac{\partial F}{\partial x'}
-\Big(k(s)y+\frac{\partial\psi}{\partial y}\Big)
\frac{\partial F}{\partial y'}=0,
\end{eqnarray}

\begin{eqnarray}
&&\Big(\frac{\partial^2}{\partial x^2}+\frac{\partial^2}{\partial y^2}\Big)
\psi=-\frac{2\pi K}{N}\int\ud x'\ud y' F,
\end{eqnarray}
where $\psi(x,y,s)$  is normalized electrostatic potential and  $F(x,y,x',y',s)$ is 
distribution function in transverse phase space $(x,y,x',y',s)$  with normalization 
\begin{eqnarray}
N=\int\ud x\ud y n,\qquad n(x,y,s)=\int\ud x'\ud y' F,
\end{eqnarray}
         where $K$  is self-field perveance which measures self-field intensity. Introducing 
self-field energy 
\begin{eqnarray}
E(s)=\frac{1}{4\pi K}\int\ud x\ud y |\partial^2\psi/\partial x^2+
\partial^2\psi/\partial y^2 |
\end{eqnarray}
we have obvious equations for root-mean-square beam radius   $R(s)$
    
\begin{eqnarray}
R(s)=<x^2+y^2>^{1/2}
\end{eqnarray}
and unnormalized beam emittance 
\begin{equation}
\varepsilon^2(s)=4(<x'^2+y'^2><x^2+y^2>-<xx'-yy'>),
\end{equation}
         which appear after averaging second-moments quantities regarding distribution 
function  $F$ [8]:
\begin{eqnarray}
\frac{\ud^2 R(s)}{\ud s^2}+\Big(k(s)R(s)-\frac{K(1+\Delta)}{2R^2(s)}\Big)R(s)&=&
\frac{\varepsilon^2(s)}{4R^3(s)}
\\
\frac{\ud\varepsilon^2(s)}{\ud s}+8R^2(s)\Big(\frac{\ud R}{\ud s}
\frac{K(1+\Delta)}{2R(s)}-\frac{\ud E(s)}{\ud s}\Big)&=&0,\nonumber
\end{eqnarray}
         where the term 
\begin{equation}
K(1+\Delta)/2
\end{equation}
         may be fixed in some interesting cases, but generally we have it only as average 
\begin{equation}
K(1+\Delta)/2=-<x\partial\psi/\partial x+y\partial\psi/\partial y>
\end{equation}
         w.r.t. distribution  $F$. Anyway, the rate equations represent reasonable reductions 
for the second-moments related quantities from the full nonlinear Vlasov-Poisson system. For 
trivial distributions Davidson e.a. [8] found additional reductions. For KV distribution (step-
function density) the second rate equation is trivial,
\begin{equation}
\varepsilon(s)={\rm const} 
\end{equation}
         and we have only one nontrivial rate equation for rms beam radius. The fixed-shape 
density profile ansatz for axisymmetric distributions also leads to similar situation: emittance 
conservation and the same envelope equation with two shifted constants only.

\subsection{Multiscale Representation}
         Accordingly to our approach which allows us to find exact solutions 
as for Vlasov-like systems as for rms-like systems 
we need not to fix particular case of distribution function 
$F(x,y,x',y',s)$. Our consideration is based on the following multiscale  $N$-mode anzatz: 
\begin{eqnarray}
F^N(x,y,x',y',s)=
\sum^{N}_{i_1,\dots,i_5=1}a_{i_1,\dots,i_5}
\bigotimes^5_{k=1}A_{i_k}(x,y,x',y',s),
\end{eqnarray}

\begin{equation}
\psi^N(x,y,s)=\sum^{N}_{j_1,j_2,j_3=1}b_{j_1,j_2,j_3}\bigotimes^3_{k=1}B_{j_k}(x,y,s).
\end{equation}
         These formulae provide multiresolution representation for variational solutions of 
our system. Each high-localized mode/harmonics  $A_j(s)$ corresponds to level  $j$ of resolution 
from the whole underlying infinite scale of spaces: 
\begin{equation}
\dots V_{-2}\subset V_{-1}\subset V_0\subset V_{1}\subset V_{2}\subset\dots,
\end{equation}
         where the closed subspace  $V_j (j\in {\bf Z})$ corresponds 
to level $j$ of resolution, or to 
scale $j$. The construction of tensor algebra 
based on the multiscale bases is considered in part I [7]. 
We will consider rate equations as the following operator equation. 
         Let  $L$, $P$, $Q$  be an arbitrary nonlinear (rational in dynamical variables) first-order 
matrix differential operators with matrix dimension   ($d=4$ in our case) corresponding to 
the system of equations, which act on some set of functions
\begin{equation}
\Psi\equiv\Psi(s)=\Big(\Psi^1(s),\dots,\Psi^d(s)\Big), \quad s \in\Omega\subset R
\end{equation}
         from 
\begin{equation}
L^2(\Omega): Q(R,s) \Psi(s)=P(R,s)\Psi(s)
\end{equation}
         or 
\begin{equation}
L\Psi\equiv L(R,s)\Psi(s)=0
\end{equation}
         where 
\begin{equation}
R\equiv R(s,\partial /\partial s, \Psi).
\end{equation}
         
         Let us consider now the N mode approximation for solution as the following 
expansion in some high-localized wavelet-like basis: 
\begin{equation}
\Psi^N(s)=\sum^N_{r=1}a^N_{r}\phi_r(s).
\end{equation} 				
         We will determine the coefficients of expansion from the following variational 
condition: 
\begin{equation}
L^N_{k}\equiv\int(L\Psi^N)\phi_k(s)\ud s=0
\end{equation}
         We have exactly $dN$  algebraic equations for $dN$  unknowns  $a_{r}$. So, variational 
approach reduced the initial problem to the problem of solution of functional equations at the 
first stage and some algebraic problems at the second stage. As a result we have the 
following reduced algebraic system of equations (RSAE) on the set of unknown coefficients 
$a_i^N$  of the expansion:
\begin{eqnarray}
H(Q_{ij},a_i^N,\alpha_I)=M(P_{ij},a_i^N,\beta_J),
\end{eqnarray}
where operators  $H$ and $M$  are algebraization of RHS and LHS of initial problem. 
$Q_{ij}$ ($P_{ij}$)  are the coefficients of LHS (RHS) of the initial system of differential equations and 
as consequence are coefficients of RSAE (39).
\begin{equation}
I=(i_1,...,i_{q+2}), \qquad J=(j_1,...,j_{p+1})
\end{equation}
         are multiindices, by which are labelled $\alpha_I$  and $\beta_I$, the other coefficients of RSAE:
\begin{equation}
\beta_J=\{\beta_{j_1...j_{p+1}}\}=\int\prod_{1\leq j_k\leq p+1}\phi_{j_k},
\end{equation}
         where $p$  is the degree of polynomial operator   
\begin{equation}
\alpha_I=\{\alpha_{i_1}...\alpha_{i_{q+2}}\}=\sum_{i_1,...,i_{q+2}}\int
\phi_{i_1}...\dot{\phi_{i_s}}...\phi_{i_{q+2}},
\end{equation}
         where  $q$ is the degree of polynomial operator $Q$, 
\begin{equation}
i_\ell=(1,...,q+2),\quad \dot{\phi_{i_s}}=\ud\phi_{i_s}/\ud s.
\end{equation}
         We may extend our approach to the case when we have additional constraints on the 
set of our dynamical variables 
\begin{equation}
\Psi=\{R, \varepsilon\}
\end{equation}
         and additional averaged terms also. In this case by using the method of Lagrangian 
multipliers we again may apply the same approach but for the extended set of variables. As a 
result we receive the expanded system of algebraic equations analogous to our system. 
Then, after reduction we again can extract from its solution the coefficients of the expansion. 
It should be noted that if we consider only truncated expansion with  $N$ terms then we have 
the system of  $N\times d$ algebraic equations with the degree
\begin{equation}
\ell=max\{p,q\}
\end{equation}
         and the degree of this algebraic system coincides with the degree of the initial 
system. So, after all we have the solution of the initial nonlinear (rational) problem in the form 
\begin{eqnarray}
R^N(s)=R(0)+\sum_{k=1}^Na_k^N \phi_k(s),\qquad
\varepsilon^N(s)&=&\varepsilon(0)+\sum_{k=1}^Nb_k^N \phi_k(s),
\end{eqnarray}
where coefficients 
 \begin{equation}
a_k^N,\qquad b_k^N
\end{equation}				
         are the roots of the corresponding reduced algebraical (polynomial) problem RSAE. 
Consequently, we have a parametrization of the solution of the initial problem by solution of 
reduced algebraic problem. The problem of computations of coefficients 
\begin{equation}
\alpha_I , \quad \beta_J 
\end{equation}
         of reduced algebraical system may be explicitly solved in wavelet approach. 
         The obtained solutions are given in the form, where  $\phi_k(s)$ are proper wavelet bases 
functions (e.g., periodic or boundary). 
         It should be noted that such representations give the best possible localization 
properties in the corresponding (phase)space/time coordinates. 
         In contrast with different approaches these formulae do not use perturbation 
technique or linearization procedures and represent dynamics via generalized nonlinear 
localized eigenmodes expansion. 
         Our  $N$ mode construction gives the following general multiscale representation:
 \begin{eqnarray}
R(s)&=&R_{N}^{slow}(s)+\sum_{i\ge N}R^i(\omega_is),
\quad \omega_i \sim 2^i,\\
\varepsilon(s)&=&\varepsilon_{N}^{slow}(s)+\sum_{j\ge N}\varepsilon^j(\omega_js),
\quad \omega_j \sim 2^j,\nonumber
\end{eqnarray}
         where $R^i(s), \varepsilon^j(s)$  are represented by some 
family of (nonlinear) eigenmodes and 
gives the full multiresolution/multiscale representation in the high-localized wavelet bases. As 
a result we can construct 
various unstable (Fig. 10, Part I) or stable patterns (Fig. 1, 2 , Part II 
and Fig. 11, Part I) from high-localized (coherent) fundamental modes (Fig. 1 or 8 or 9 from 
Part I) in complicated stochastic systems with complex collective behaviour. Definitely, 
partition(s) as generic dynamical variables cannot be postulated but need to be computed as 
solutions of proper stochastic dynamical evolution model. Only after that, it is possible to 
calculate other dynamical quantities and physically interesting averages.


\

\section{CONCLUSIONS: TOWARDS ENERGY\\ CONFINEMENT}

         Analysis and modeling considered in these two Parts describes, in principle, a 
scenario for the generation of controllable localized (meta) stable 
fusion-like state (Fig. 2 and 
Fig. 6). Definitely, chaotic-like unstable partitions/states dominate during non-equilibrium 
evolution. It means that (possible) localized (meta) stable partitions have measure equal to 
zero a.e. on the full space of hierarchy of partitions defined on a domain of the definition in 
the whole phase space. Nevertheless, our Generalized Dispersion Relations, like (15) or (39), 
give some chance to build the controllable 
localized state (Fig. 6) starting from initial chaotic-like 
partition (Fig. 3) via process of controllable self-organization.
         Figures 4 and 5 demonstrate two subsequent steps towards creation important fusion 
or confinement state, Fig. 6, which can be characterized by the presence of a few important 
physical modes only in contrast with the opposite, chaotic-like state, Fig. 3, described by infinite 
number of important modes. Of course, such confinement states, characterized by zero 
measure and minimum entropy, can be only metastable. But these long-living fluctuations can 
and must be very important from the practical point of view, because the averaged time of 
existence of such states may be even more than needed for practical realization, e.g., in 
controllable fusion processes.
         Further details will be considered elsewhere.

\section*{ACKNOWLEDGEMENTS}
         We are very grateful to Professors E. Panarella (Chairman of the Steering 
Committee), R. Kirkpatrick (LANL) and R.E.H. Clark, G. Mank and his Colleagues from 
IAEA (Vienna) for their help, support and kind attention before, during and after 6th 
Symposium on Fusion Research (March 2005, Washington, D.C.). We are grateful to Dr. 
A.G. Sergeyev for his permanent support in problems related to hard- and software.

\newpage
\begin{figure}
\centering 
\includegraphics[width=90mm]{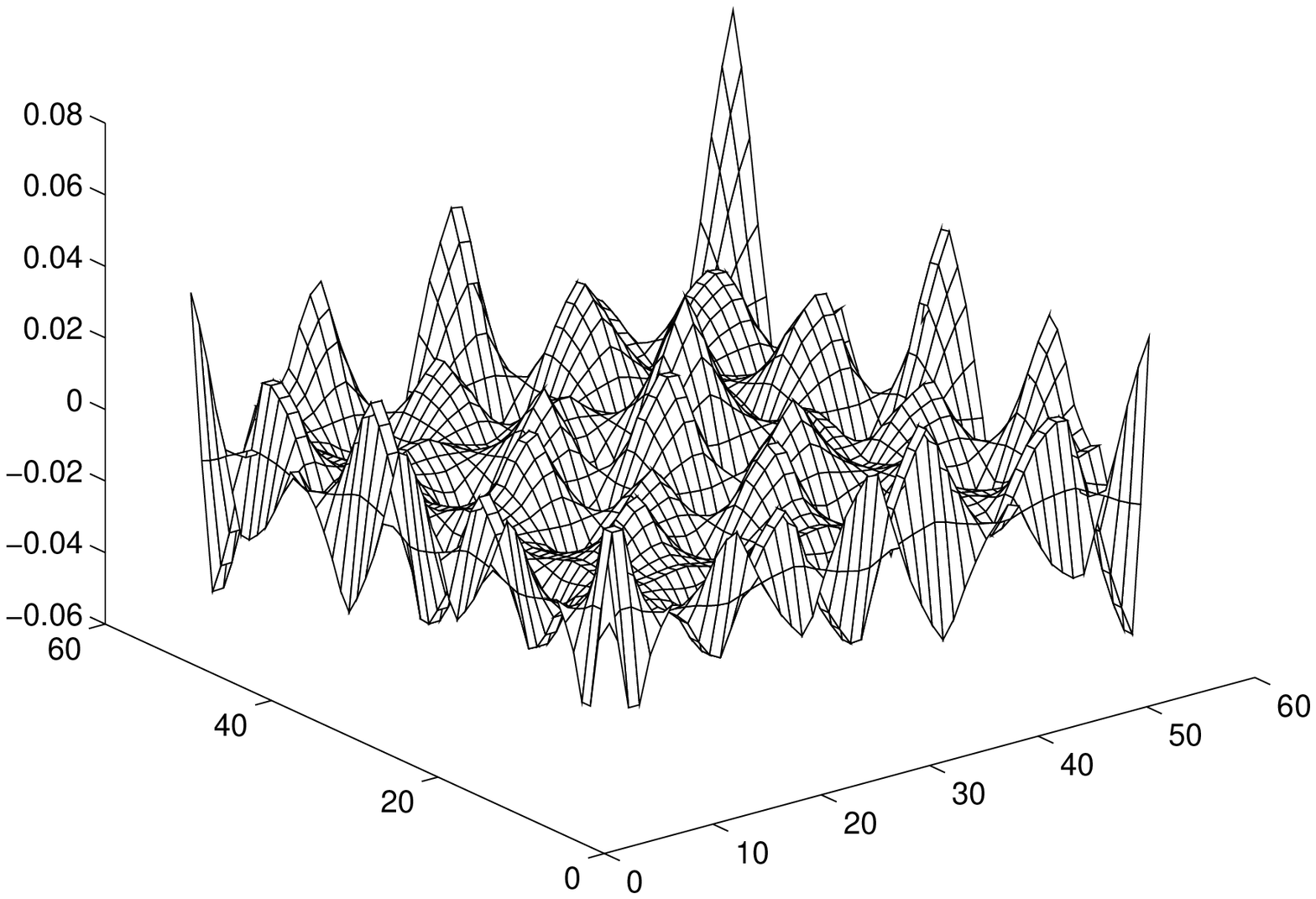} 
\caption{Localized two-dimensional partition.}
\end{figure}   

\begin{figure}
\centering 
\includegraphics[width=90mm]{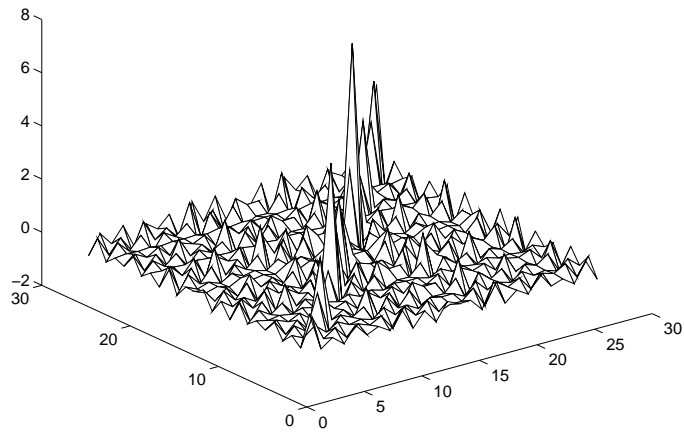} 
\caption{Metastable pattern.}
\end{figure}

\begin{figure}[htb]
\centering 
\includegraphics[width=90mm]{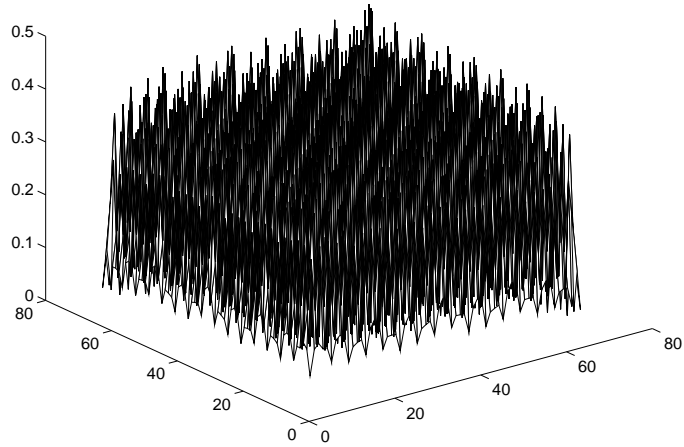} 
\caption{Chaotic pattern.}
\end{figure}

\begin{figure}[htb]
\centering 
\includegraphics[width=90mm]{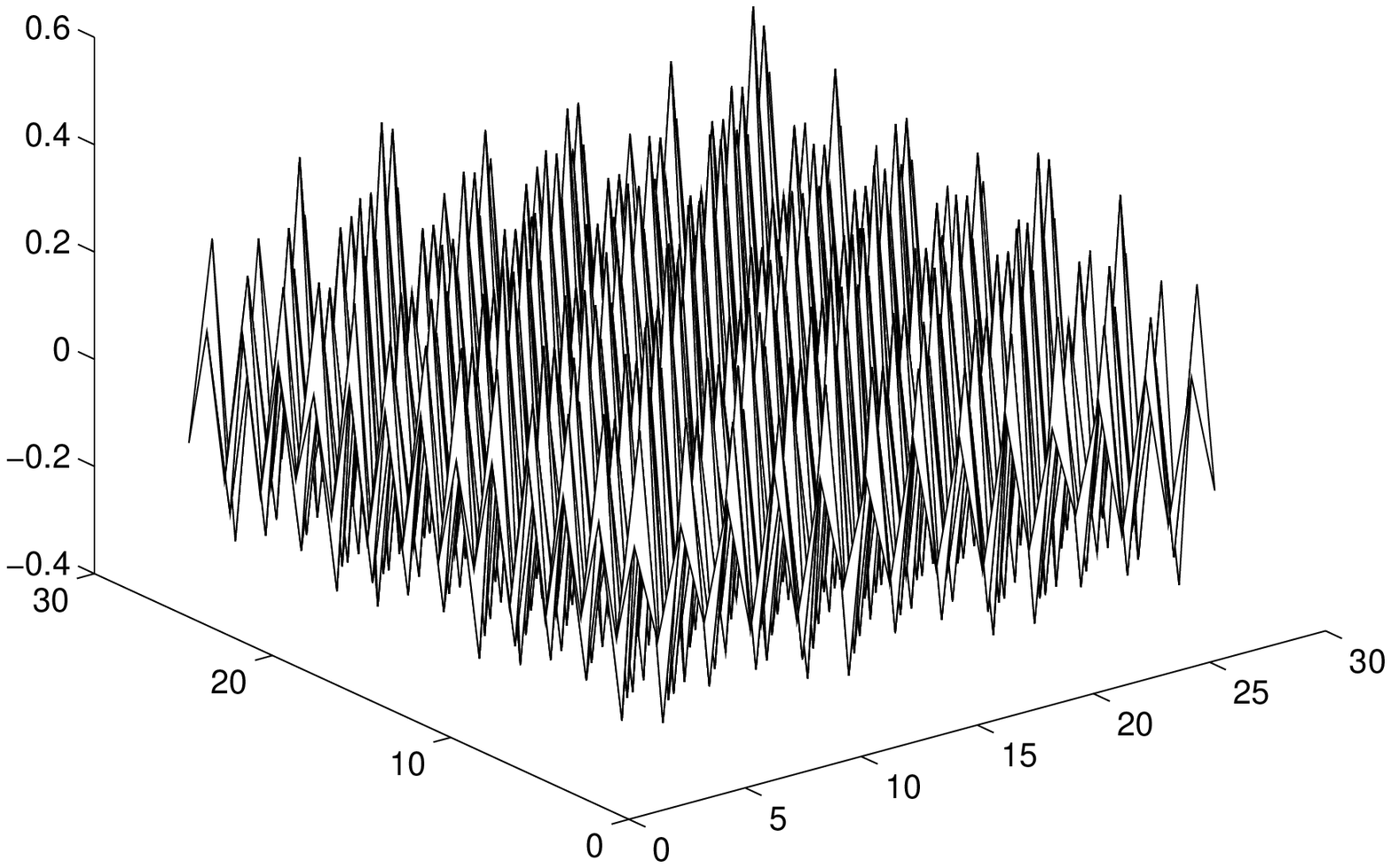} 
\caption{Self-organization, step 1: towards confinement.}
\end{figure}  

\newpage
\begin{figure}[htb]
\centering 
\includegraphics[width=90mm]{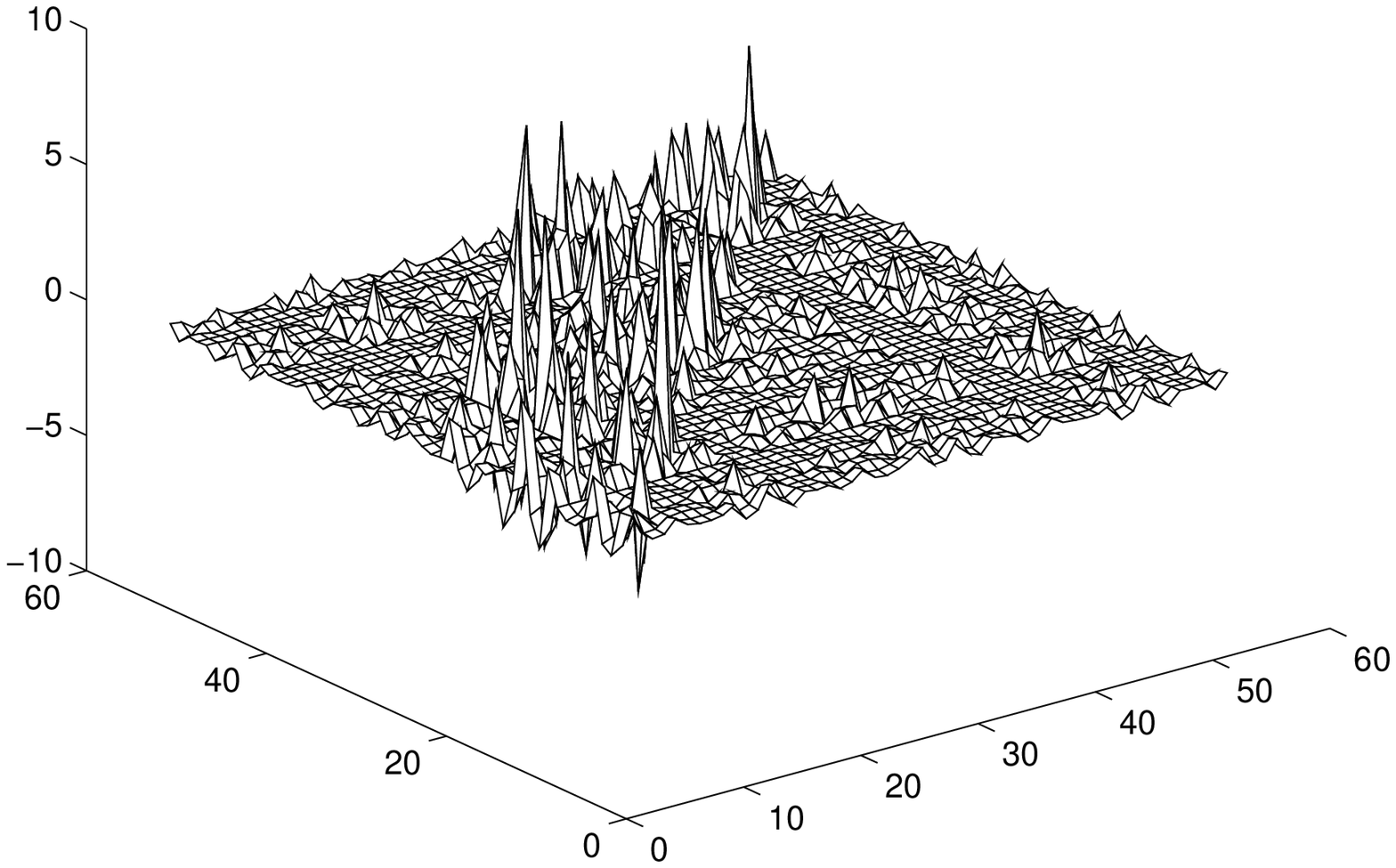} 
\caption{Self-organization, step 2: towards confinement.}
\end{figure} 

\begin{figure}[htb]
\centering 
\includegraphics[width=90mm]{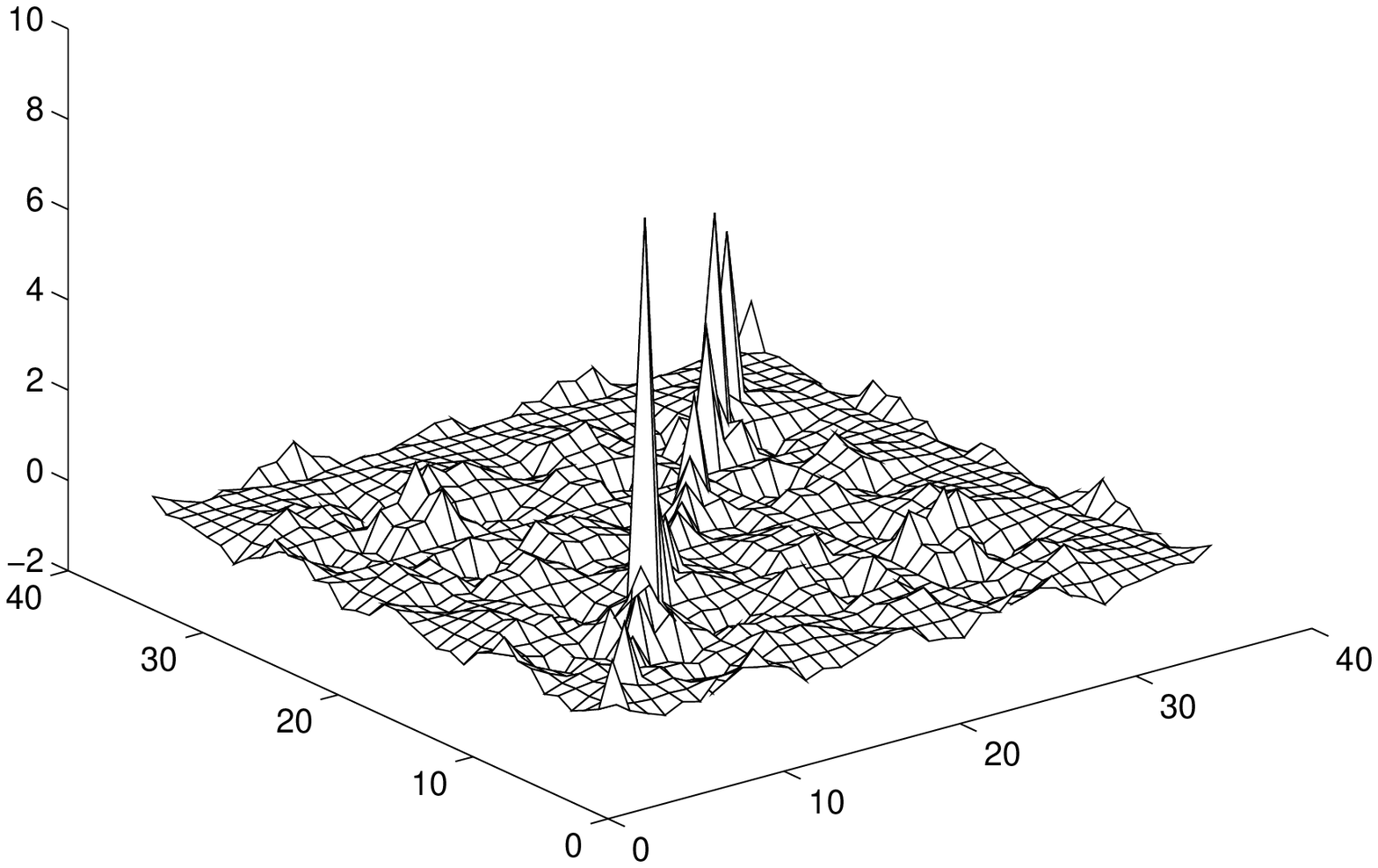} 
\caption{Localized pattern(waveleton): energy confinement state.}
\end{figure}


\begin{thebibliography}{9}

\bibitem{1}	A.N. Fedorova and M.G. Zeitlin, Math. and Comp. in Simulation, 46, 527 (1998); New 
Applications of Nonlinear and Chaotic Dynamics in Mechanics, Ed. F. Moon, (Kluwer, 
Boston, 1998) pp. 31-40, 101-108. 

\bibitem{2}	A.N. Fedorova and M.G. Zeitlin, in American Institute of Physics, Conf. Proc. 405 
(1997), pp. 87-102; "Nonlinear Dynamics of Accelerator via Wavelet Approach", 
physics/9710035; 468 (1999), pp. 48-68, 69-93; "Variational Approach in Wavelet 
Framework to Polynomial Approximations of Nonlinear Accelerator Problems", 
physics/990262; "Symmetry, Hamiltonian Problems and Wavelets in Accelerator 
Physics", physics/990263.
 
\bibitem{3}	A.N. Fedorova and M.G. Zeitlin, in The Physics of High Brightness Beams, Ed. J. 
Rosenzweig, 235, (World Scientific, Singapore, 2001) pp. 235-254; "Variational-
Wavelet Approach to RMS Envelope Equations", physics/0003095. 

\bibitem{4}	A.N. Fedorova and M.G. Zeitlin, in Quantum Aspects of Beam Physics, Ed. P. Chen 
(World Scientific, Singapore, 2002) pp. 527-538, 539-550; "Quasiclassical Calculations 
for Wigner Functions via Multiresolution", physics/0101006; "Localized Coherent 
Structures and Patterns Formation in Collective Models of Beam Motion", 
physics/0101007. 

\bibitem{5}	A.N. Fedorova and M.G. Zeitlin, in Progress in Nonequilibrium Green's Functions II, 
Ed. M. Bonitz, (World Scientific, 2003) pp. 481-492; "BBGKY Dynamics: from 
Localization to Pattern Formation", physics/0212066. 

\bibitem{6}	A.N. Fedorova and M.G. Zeitlin, in Quantum Aspects of Beam Physics, Eds. Pisin Chen, 
K. Reil (World Scientific, 2004) pp. 22-35; "Pattern Formation in Wigner-like Equations 
via Multiresolution", SLAC-R-630 and quant-phys/0306197; Nuclear Instruments and 
Methods in Physics Research Section A, 534, Issues 1-2, 309-313, 314 -318 (2004); 
"Classical and Quantum Ensembles via Multiresolution. I. BBGKY Hierarchy", quant-
ph/0406009; "Classical and Quantum Ensembles via Multiresolution. II. Wigner 
Ensembles", quant-ph/0406010.
 
\bibitem{7}	A.N. Fedorova and M.G. Zeitlin, Localization and Fusion Modeling in Plasma Physics. 
Part I: Math Framework for Non-Equilibrium Hierarchies, this volume.

\bibitem{8}	R.C. Davidson and H. Qin, Physics of Intense Charged Particle Beams in High Energy 
Accelerators (World Scientific, Singapore, 2001); A.W. Chao, Physics of Collective 
Beam Instabilities in High Energy Accelerators (Wiley, New York, 1993). R. Balescu, 
Equilibrium and Nonequilibrium Statistical Mechanics, (Wiley, New York, 1975); C. 
Scovel, A. Weinstein, Comm. Pure. Appl. Math., 47, 683, 1994; H. Boozer, Rev. Mod. 
Phys., 76, 1071 (2004). 
\bibitem{9}	Y. Meyer, Wavelets and Operators (Cambridge Univ. Press, 1990); D. Donoho, WaveLab 
(Stanford, 2000). 
\end{thebibliography}
\end{document}